# A Three-terminal Non-Volatile Ferroelectric Switch with an Insulator-Metal Transition Channel

Jaykumar Vaidya[1,a], R S Surya Kanthi[1,a], Shamiul Alam[2], Nazmul Amin[2], Ahmedullah Aziz[2], Nikhil Shukla[1]*

[1]Department of Electrical and Computer Engineering, University of Virginia, Charlottesville, VA- 22904, USA

[2]Department of Electrical Engineering and Computer Science, University of Tennessee, Knoxville, TN 37996, USA

a) equal contribution

*e-mail: ns6pf@virginia.edu

## Abstract

Ferroelectrics offer a promising materials platform to realize energy-efficient non-volatile memory technology with the FeFET-based implementations being one of the most area-efficient ferroelectric memory architectures. However, the FeFET operation entails a fundamental trade-off between the read and the program operations. To overcome this trade-off, we propose in this work, a novel device, Mott-FeFET, that aims to replace the Silicon channel of the FeFET with $VO_2$- a material that exhibits an electrically driven insulator-metal phase transition. The Mott-FeFET design, which demonstrates a (ferroelectric) polarization-dependent threshold voltage, enables the read current distinguishability (i.e., the ratio of current sensed when the Mott-FeFET is in state 1 and 0, respectively) to be independent of the program voltage. This enables the device to be programmed at low voltages without affecting the ability to sense/read the state of the device. Our work provides a pathway to realize low-voltage and energy-efficient non-volatile memory solutions.

The electric-field (E-field) induced non-volatile polarization switching in ferroelectrics makes them a promising candidate for developing non-volatile memory (NVM) technology. Conventionally, ferroelectric-based random-access memory (RAM) was realized using traditional ferroelectrics such as PZT, and showed energy-efficient operation, fast read as well as high endurance[1,2]. However, this ferroelectric memory technology was challenging to scale since ferroelectrics such as PZT exhibit a significant degradation in the ferroelectric response when the film thickness is scaled below 50nm[3]. Consequently, the recent discovery of ferroelectricity in highly scaled $HfO_2$- a material that is compatible with CMOS process technology- has generated immense interest in revisiting ferroelectric memory technology[4,5]. Particularly, the ability to integrate the ferroelectric directly into the gate of a field effect transistor (FET) has motivated active investigation of 1T-FeFET (1 Transistor-Ferroelectric Field Effect Transistor)-based non-volatile memory[6]. While the 1T-FeFET architecture facilitates the realization of a dense array, it entails a fundamental trade-off between the programming and the read/sensing characteristics[7–9]. The objective of this work is to propose a pathway to overcome this trade-off by replacing the Silicon channel by an alternate channel material, $VO_2$ (vanadium dioxide), that exhibits the phenomenon of electrically driven insulator-to-metal transition (IMT).

A conventional FeFET involves a fundamental trade-off between the program voltage (write operation), and the MW (memory window) along with the corresponding read current distinguishability, expressed as $I_{bit\_1}/I_{bit\_0}$ ($I_{bit\_1}$ and $I_{bit\_0}$ are the sense currents measured corresponding to bit 1 and 0, respectively). Increasing the memory window and the corresponding $I_{bit\_1}/I_{bit\_0}$ requires the application of a significantly larger programming voltage. This is because in the FeFET configuration, the ferroelectric typically operates on a minor loop (not saturation loop) of the polarization vs. voltage characteristics and improving the MW entails increasing the hysteresis by the application of a larger programming voltage. Moreover, these contending factors can become even more critical while operating the cell in a memory array where the parasitic currents from half-selected cells can further compromise the read distinguishability. Additionally, the larger program voltage also results in extremely large electric-fields (in excess of 10 MV/cm) across the interlayer (IL) between the ferroelectric and the Silicon channel which can adversely impact the reliability and the endurance of the device[8]. These trade-offs have been quantitatively analyzed in prior works[7,9] including those by the authors[8,10].

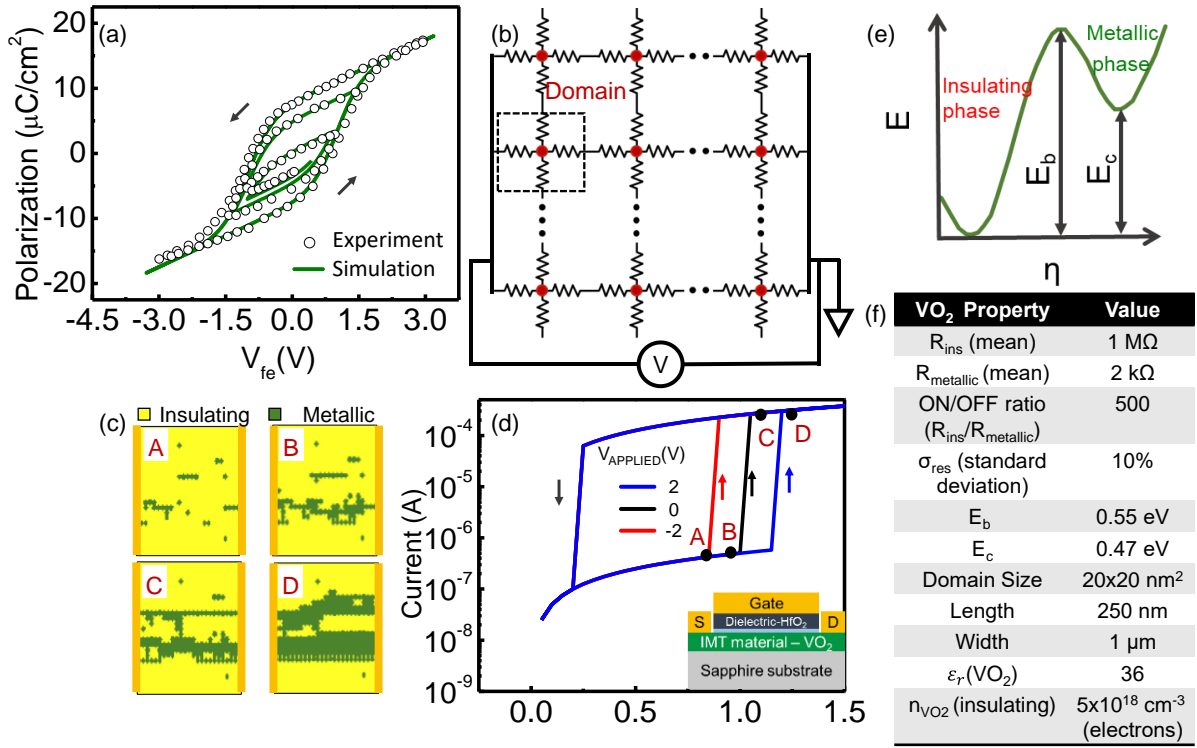

**Fig. 1. Modelling electrical response of the ferroelectric and VO₂.** (a) Polarization vs. voltage characteristics of the ferroelectric HfO$_2$ simulated using the Preisach's model, and calibrated to the experimental data reported by S. Mueller et al.[12] (b) Two-dimensional resistive network used to model the filamentary switching behavior across the IMT in VO$_2$. (c) Evolution of the insulating and metallic phases across the electrically driven IMT in VO$_2$ clearly showing the filamentary switching; the phases are indicated on the black IV curve in next panel (d) Modulation of the trigger voltage for the IMT as a function of the applied electric field. Schematic of the device considered is also shown (inset) (e) Phase diagram of VO$_2$ (f) Parameters used for simulation of VO$_2$.

To illustrate the concept of the Mott-FeFET, we first develop models for the individual components of the device, namely, the ferroelectric HfO$_2$ and the VO$_2$ channel. Fig. 1a shows the simulated polarization vs. voltage characteristics of the ferroelectric HfO$_2$ (*f*-HfO$_2$) considered in this work.

These characteristics have been simulated using the phenomenological Preisach's model[11] and have been calibrated to experimental data on 10nm thick *f*-HfO$_2$ films (in a metal/ferroelectric/metal capacitor configuration) reported by S. Mueller *et al.*[12].

Next, we consider the electronically driven IMT in VO$_2$. We note that while other oxides such as NbO$_2$[13], SmNiO$_3$[14] among others also demonstrate similar electronically driven IMT behavior, VO$_2$ was our choice of IMT material in this work since it exhibits a large R$_{OFF}$/R$_{ON}$ ratio (>10$^4$)[15], and more importantly, the modulation of the threshold voltage of the electrical IMT with gate electric field, critical to the Mott-FeFET operation, has been experimentally demonstrated in this material system[16,17].

Two-terminal VO$_2$ devices exhibit an electrically induced IMT that is characterized by an abrupt change in resistance at a particular applied voltage (V$_T$: threshold voltage) as the device transitions from the insulating to the metallic state[18]. The transition is volatile, and the device returns back to the insulating state (metal-to-insulator transition; MIT) when the applied voltage subsequently drops below a threshold (V$_H$: hold voltage), accompanied by hysteresis (V$_T$-V$_H$). To simulate the electronically induced IMT in VO$_2$, we model the two-terminal VO$_2$ device as a network of resistors that represent domains (Fig. 1b), as proposed in prior work[19]. Each resistance in this 2D network can undergo an IMT and MIT (metal to insulator transition) with a certain probability that is dependent on the voltage. Using the approach proposed by Madan *et al.*[15] & Poklonski *et al.*[20], we model the switching probability for a domain using the following equations:

$$P_{IMT} = e^{\frac{-\left(E_b - \frac{q\Delta V}{\gamma}\right)}{kT}} \qquad (1)$$

And,

$$P_{MIT} = e^{-\frac{(E_b - E_c)}{kT}} \qquad (2)$$

where $P_{MIT}$ and $P_{IMT}$ are the probabilities of a domain undergoing MIT and IMT, respectively; $E_b$ is the energy barrier between the insulating and the metallic state, $\gamma$ is a geometric factor[15], $\psi_s$ is the VO$_2$ surface potential due to the gate and $\alpha$ is the coupling constant introduced between the gate-induced surface potential and the IMT transition in the VO$_2$ device (set to 0.5). $E_b$ and $E_c$ are defined in Fig. 1e.

It can be observed that applying a voltage increases the probability of a domain undergoing IMT. Additionally, we also consider a gaussian distribution for the resistance values representing a domain to account for the heterogeneity in the film. The parameters for the IMT in VO$_2$ are shown in the table in Fig. 1f. A detailed discussion of the electronic IMT in VO$_2$ has been included in supplement S3.

The voltage-induced IMT in the device can be explained as follows. Initially, all the domains are in the insulating state (at zero bias). As the voltage across the device is increased, a few domains (probabilistically) undergo IMT, serving as the nucleation centers for the metallic phase. As the domain transitions to the metallic state, the voltage drop across the domain reduces, leading to a corresponding increase in the voltage drop across other domains, which in turn, increases their probability of switching. This process generates an avalanching effect that eventually creates a metallic filamentary bridge between the electrodes, resulting in an abrupt change in resistance of the device; the width of the filament on the current passing through the device- an effect that is captured by our model as well (Fig. 1c). Furthermore, the presence of filamentary conduction has been experimentally shown in prior work[21]. Similar (albeit weaker) avalanching behavior is observed during MIT leading to an abrupt increase in resistance as the device turns OFF. We also note that since the switching in VO$_2$ is stochastic (see supplement S1a) which consequently has important implications for the design of the memory array.

Additionally, in the three-terminal device configuration with a gate dielectric, Kim *et al.*[16], and Tabib-Azar *et al.*[17] experimentally demonstrated that the threshold voltage of the VO$_2$ channel can be modulated by applying an electric field across the gate – a property crucial to the Mott-FeFET operation. We note that the gate-field alone does not induce the IMT but aids the transition. We model this behavior *phenomenologically* by modifying equation (1) to include the effect of surface potential induced by the gate: $P_{IMT} = e^{\frac{-\left(E_b - \frac{q\Delta V}{\gamma} - \alpha q \psi_s\right)}{kT}}$; the surface potential of VO$_2$ (which is calculated using a capacitance divider analysis, similar to the gate stack in a standard FET) is coupled to the probability of the IMT through a coupling constant α. Since the gate-field alone does not induce the transition but modifies the threshold voltage ($V_T$), we model this effect as the surface potential modulating the probability of the domain switching, which subsequently, manifests as the change in the (drain-to-source) $V_T$ required to induce the IMT, as shown in Fig. 1d. Additionally, we do not consider this effect in the MIT characteristics since the operation of our proposed device as a memory cell does not rely on the MIT, as well as due to the absence of experimental data. We also emphasize that while the proposed model can explain the experimental behavior shown by Kim *et al.*[16] *and* Tabib-Azar *et al.*[17], it is important to qualify that model is phenomenological in nature; the exact physics of the electrically induced Mott-Peierls IMT in VO$_2$ still remains an active, ongoing investigation.

Another important aspect of the gate electric-field induced modulation is that while it has a significant influence on the threshold voltage for the IMT (by influencing the nucleation dynamics of the metallic phase), its impact on the metallic and insulating states is minimal. This because the high conductance state of the VO$_2$ is essentially metallic in nature which limits the penetration of the gate field. Furthermore, the impact of the electric field on the resistivity of insulating state is also expected to be minimal[22]. This can be attributed to the formation of small polarons that result from the gate-induced charge coupling to the lattice, as shown in our prior work[23], as well as in other works[24]. These polarons screen the electric field, and subsequently, limit its penetration to a few (1-2) monolayers, resulting in minimal effect on the conductivity. This ensures that for a current sensing-based reading scheme, the read distinguishability i.e., $I_{bit\_1}$/ $I_{bit\_0}$ would essentially be constant, irrespective of the programming voltage / field applied at the gate. This behavior is fundamentally different from that of a conventional Si transistor where the gate field strongly controls the channel resistance, and consequently, the channel current.

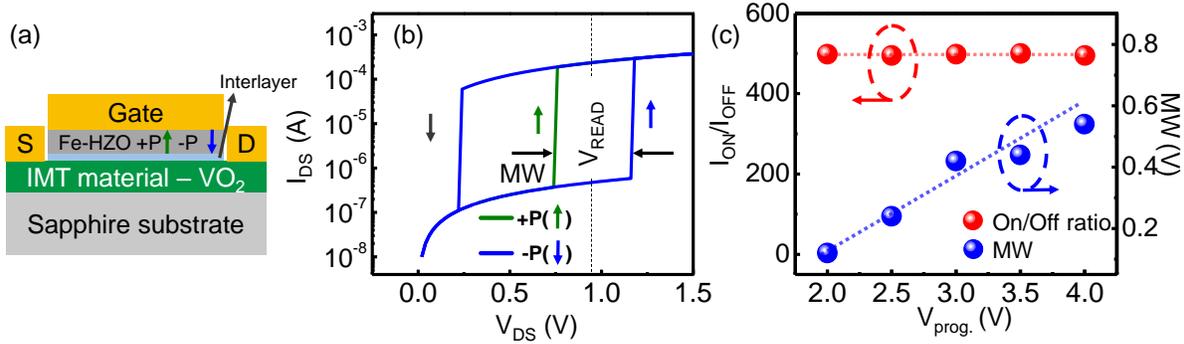

**Fig. 2. Mott-FeFET operation.** (a) Schematic of the proposed Mott-FeFET. (b) $I_{ds}$ vs. $V_{ds}$ characteristics of the $VO_2$ channel as a function of the ferroelectric polarization. (c) Observed ratio between read currents corresponding to state 1 and 0, as a function of the applied program voltage. It can be observed that $I_{ON}/I_{OFF}$ ratio, unlike in a conventional FeFET remains constant.

The Mott-FeFET design (Fig. 2a) aims to integrate the non-volatile polarization switching in the ferroelectric gate with the abrupt resistance switching across the electrically driven IMT in $VO_2$ channel. The expected operation of the Mott-FeFET can be described as follows: the polarization state (up or down) of the ferroelectric gate is used to represent the information bit to be stored and can be programmed (write operation) by applying the program voltage across the gate of the device (details of the polarity are discussed in the array operation). The resulting surface potential associated with the (different) polarization states of the ferroelectric modulates the threshold voltage ($V_{T,1}$, $V_{T,0}$) of the IMT in the $VO_2$ channel i.e., one polarization state results in a larger threshold voltage than the other state, creating a memory window ($\Delta V_T = V_{T,0} - V_{T,1}$) as shown in Fig. 2b. Subsequently, the state of the memory can be sensed by applying an appropriate read voltage $V_{READ}$ such that $V_{T,1} < V_{READ} < V_{T,0}$. This ensures that if the memory cell is in state 1, a large drain current, corresponding to the metallic state of $VO_2$, will be sensed whereas state 0 will produce a significantly smaller drain current owing to the insulating nature of the channel.

Therefore, the VO$_2$ channel can be considered as a 'selector' whose threshold voltage depends on the state of the memory (i.e., ferroelectric gate).

We explore the operation of the proposed Mott-FeFET by integrating the models developed above for the individual components, namely, polarization switching in the ferroelectric and the electronically induced abrupt resistance switching in VO$_2$. Moreover, the ferroelectric polarization interacts with the VO$_2$ channel through the surface potential which is calculated by modeling the capacitance response of the gate stack. The surface potential, which depends on the state of polarization of the ferroelectric, subsequently, modulates the probability of the switching (from insulating to metallic state) in the VO$_2$ domains, resulting in a ferroelectric polarization state dependent IMT threshold voltage. A finite interlayer at the interface between the ferroelectric and the VO$_2$ is also considered. Using this framework, we simulate the characteristics of the Mott-FeFET, as shown in Fig. 2b. It can be observed that the threshold voltage for the IMT in the VO$_2$ channel varies by ~0.5V, opening a memory window that can facilitate its use a non-volatile storage element.

We note that while the memory window (i.e., $\Delta V_T = V_{T,0} - V_{T,1}$) is sensitive to the polarization, which in turn depends on the voltage used to program the ferroelectric, the $I_{bit\_1}/I_{bit\_0}$ ratio (within the memory window) is almost insensitive to the program voltage of the ferroelectric, as observed in Fig. 2c. This is because the current distinguishability is primarily decided by the $R_{OFF}$ and $R_{ON}$ of the VO$_2$ which are insensitive to the gate-field, as discussed above. The insensitivity of the $I_{bit\_1}/I_{bit\_0}$ ratio to the program voltage can facilitate scaling of the program voltage without adversely impacting the read/sense characteristics and margins.

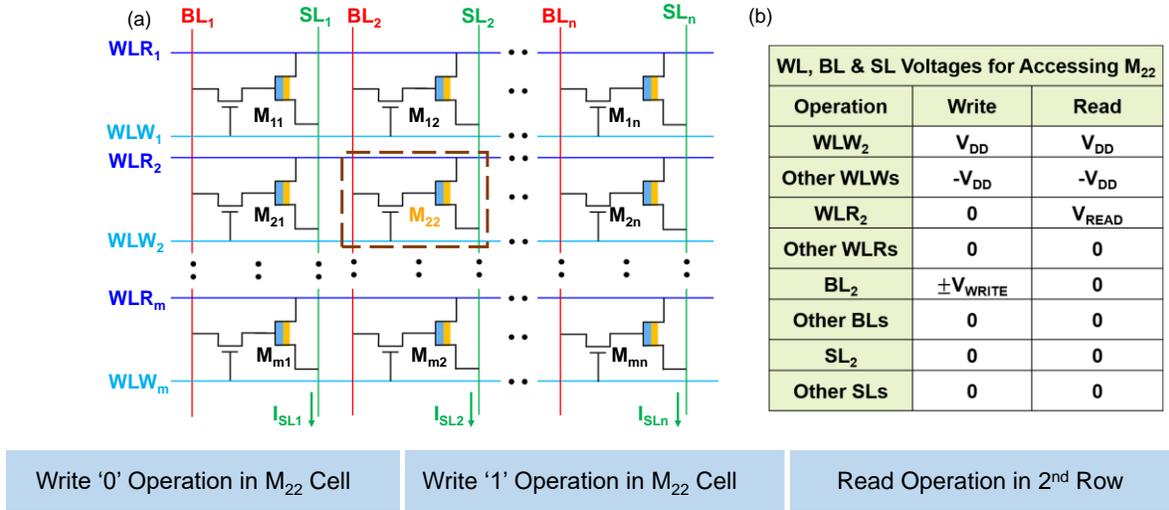

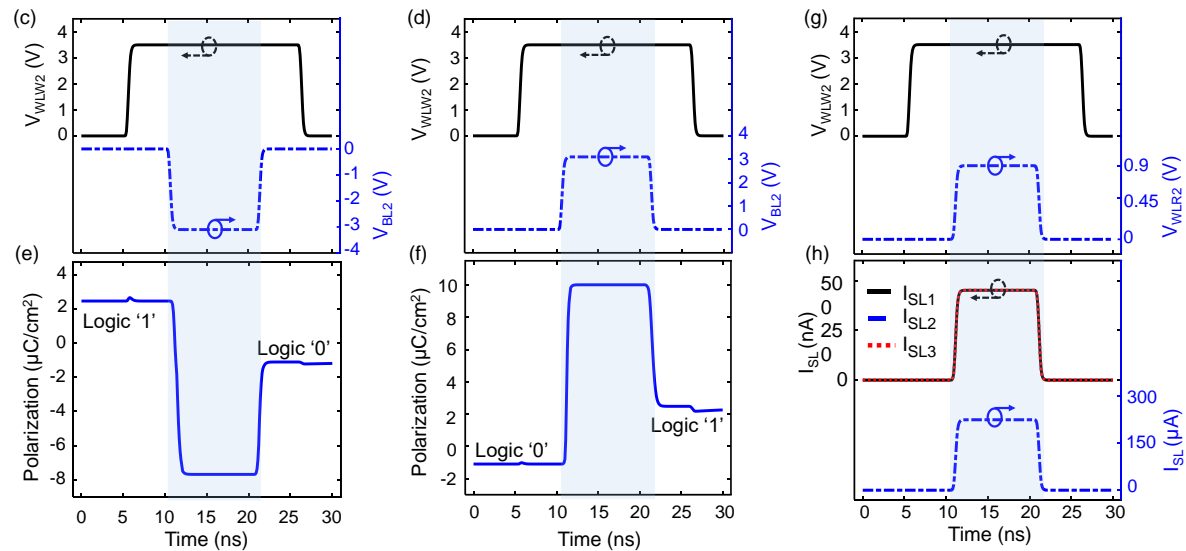

**Figure 3. Mott-FeFET array operation.** (a) Schematic of the proposed Mott-FeFET-based memory array. (b) Biasing scheme for WLWs, WLRs, BLs and SLs to access a memory cell ($M_{22}$ here). The results presented here are for a 3×3 array. Time dynamics of the bias voltages applied across WLW2 and BL2 during (c) write '0' and (d) write '1' operations in the $M_{22}$ cell. During write operation, the other WLWs and BLs are biased at $V_{DD}$ and 0 V, respectively, and all the WLRs and SLs are biased at 0 V. Time evolution of the ferroelectric polarization for (e) write '1' → '0' and (f) write '0' → '1' operations in the $M_{22}$ cell. (g) Temporal dynamics of the bias voltages of WLW2 and WLR2 for read operation of the cells in the second row. The bias voltages for other WLWs, WLRs, BLs and SLs are kept constant at specific levels (shown in (b)). (h) SL currents during read operation. This array architecture facilitates reading the entire row in one cycle. Here, we only read the second row. $M_{21}$ and $M_{23}$ cells were initialized with logic '0' and $M_{22}$ with logic '1' before read operation. The effect of the stored memory state is observed in the corresponding SL currents. $SL_1$ and $SL_3$ currents are 450 nA due to logic '0' in $M_{21}$ and $M_{23}$ and $SL_2$ current is 225 μA due to logic '1' stored in $M_{22}$.

Next, we evaluate the operation of the Mott-FeFET as a memory element in a non-volatile memory array. We consider the NOR memory architecture as shown in Fig. 3a, where the basic building block of the array consists of the Mott-FeFET as the memory element whose gate is connected to a simple MOSFET, which functions as the access device. This architecture, which consists of a separate word-line to read (WLR) from, and write (WLW) to, a cell is similar to that proposed for FeFET-based memory arrays[25]. A phenomenological Verilog-A model is used to simulate the Mott-FeFET whereas the DGXFET NMOS model, available in the IBM 65nm CMOS 10LPe process, is used for the transistor.

The biasing scheme for reading from-, and writing to-, a particular cell of the memory array is designed to facilitate successful reading and writing operations, without disturbing other cells in the array. Here, we consider the illustrative example of accessing the $M_{22}$ cell in a 3x3 array. The corresponding biasing conditions are shown in Fig. 3b. Figures 3c,d show the bias voltages applied to $WLW_2$ and $BL_2$ (connected with $M_{22}$) for the write '0' and write '1' operations, respectively. The bias conditions for the other WLWs, BLs, WLRs, and SLs are in accordance with those listed in the table in Fig. 3b. The $WLW_2$ is asserted to turn ON the access transistors of the second row, and a suitable programming voltage ($\pm V_{WRITE}$) is applied to the $BL_2$ with the objective to facilitate sufficient bias at the gate of the Fe to modulate the polarization, as needed. Figures 3e,f show the evolution of the ferroelectric polarization during write '1' → '0' and write '0' → '1' operations, respectively. The choice of the suitable bias conditions eliminates the possibility of the accidental write into the other cells of the array (details discussed in supplement S4).

The proposed array architecture also facilitates reading all the cells in a row in one cycle. To illustrate this, we initially store '0' in $M_{21}$ and $M_{23}$ and '1' in $M_{22}$ cells belonging to the second row of the 3 x 3 array. To read from a cell, we utilize the $I_{DS}$-$V_{DS}$ characteristics of Mott-FeFET (at zero gate bias) shown in Fig. 2b. Figure 3g shows the bias conditions for the $WLW_2$ and $WLR_2$ and the corresponding SL currents are shown in Fig. 3h. It can be observed that the SL connected to the cell with logic '0' generates ~450 nA whereas the logic 1 produces a current of ~225 µA on SL.

This difference in the SL current is used for the sensing of the stored memory states using current sense amplifiers[26,27] (see supplement S4 for more details on the sensing mechanism).

**Discussion**

The goal of this work is to propose and elucidate a new device concept, Mott-FeFET, that aims to overcome the read-write trade-offs in conventional Silicon FeFET designs by leveraging the unique properties of IMT. It showcases an example of how novel functional materials and their properties (here, the IMT in $VO_2$) can be used to overcome the design challenges of Silicon devices. While the focus of the work is primarily to describe the operational characteristics and functional properties of the Mott-FeFET, it is important to note that the physical realization of such a device would inevitably need to address important challenges such as the integration of the *f*-$HfO_2$ on $VO_2$ while retaining their functional properties, the role of the interfacial layer and interface states among others; overcoming these concerns will be critical to the eventual practicality of such a device. Additionally, we also note that the underlying physics of the electrically induced IMT in $VO_2$ as well as how an external electric field affects the IMT still remain to be fully understood. However, the present work helps nucleate the new device concept, and motivates the investigation of the above questions, which can subsequently, enable energy-efficient and high performance non-volatile random-access memory.

**Acknowledgments**

This work was supported by NSF grant 1914730.

**Author contributions**

J.V., R.S.S.K. performed the device simulations. S.A. and N.A performed the array simulations. J.V. and N.S. conceived the idea. A.A. and N. S. supervised the study. All authors took part in writing the manuscript, discussed the results and commented on the manuscript.

**Competing interests**

The authors declare no competing interests.

**Additional information**

Correspondence should be addressed to N. S.

**Supplementary Information**

# A Three-terminal Non-Volatile Ferroelectric Switch with an Insulator-Metal Transition Channel

Jaykumar Vaidya[1,a)], R S Surya Kanthi[1,a)], Shamiul Alam[2], Nazmul Amin[2], Ahmedullah Aziz[2], Nikhil Shukla[1]*

[1]Department of Electrical and Computer Engineering, University of Virginia, Charlottesville, VA- 22904, USA

[2]Department of Electrical Engineering and Computer Science, University of Tennessee, Knoxville, TN 37996, USA

a) equal contribution

*e-mail: ns6pf@virginia.edu

**S1. Stochastic switching in VO₂**

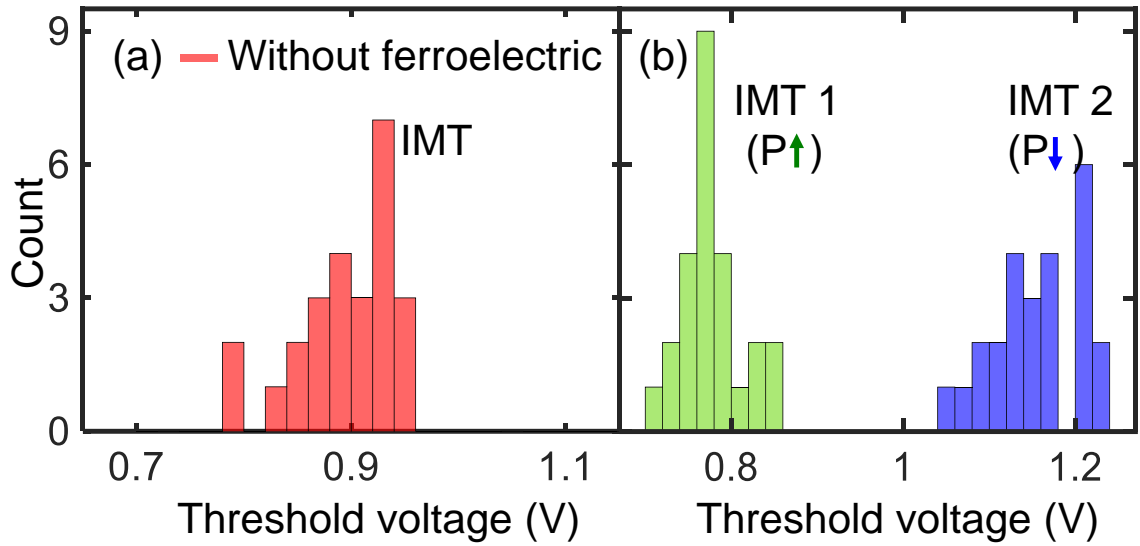

**Fig. S1. Stochastic nature of switching in VO$_2$.** (a) Distribution of IMT threshold voltage in 2 terminal VO$_2$ device (25 sweeps were considered). (b) Distribution of threshold voltages corresponding to the two states of the ferroelectric in the Mott-FeFET (25 sweeps were simulated).

## S2. Electrically driven IMT in VO₂ (Experimental characteristics)

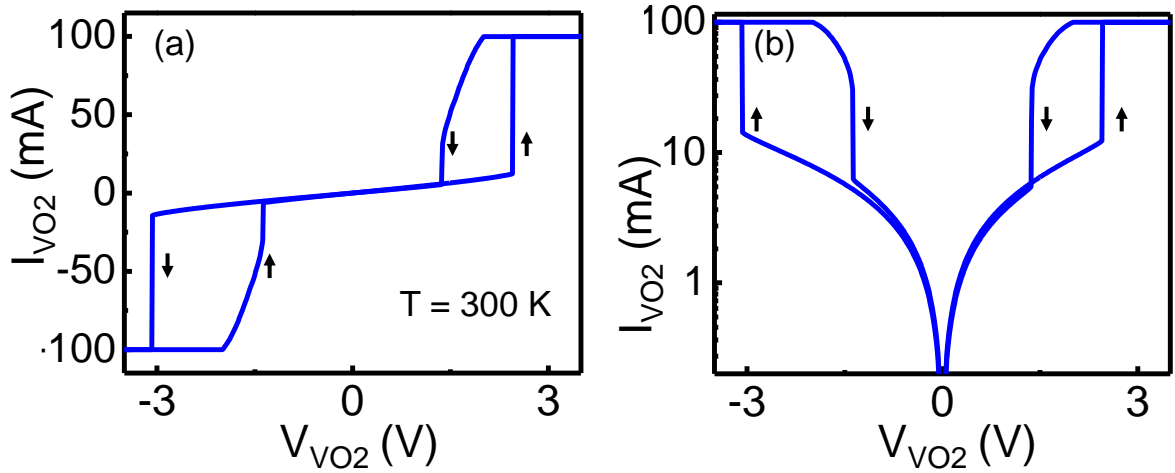

**Fig. S2. Current vs. voltage characteristics of VO$_2$.** Illustrative current vs. voltage characteristics of VO$_2$ measured experimentally in a two-terminal device configuration shown using a (a) Linear scale, (b) Logarithmic scale for the current axis.

Illustrative I-V characteristics showing electrically induced IMT and MIT in VO$_2$ marked by an abrupt and hysteretic change in resistance. The device length and width was 10 µm.

## S3. Nature of the electrical IMT in VO$_2$

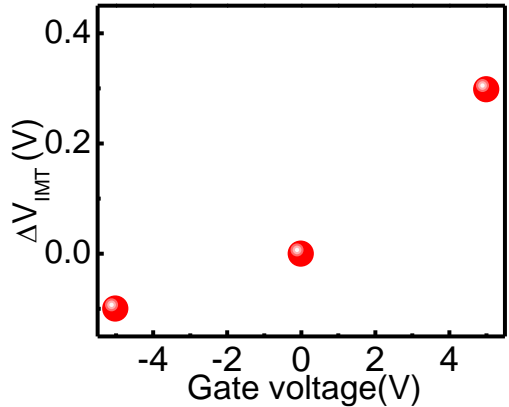

**Fig. S3** Variation in IMT threshold voltage as a function of the applied gate voltage as reported by Tabib-Azar *et al* [3].

The origin of the IMT in VO$_2$ has been the subject of intense research and debate[1]. Various models based on varying levels of contribution from a Mott-Hubbard type transition and a Peierls-like structural instability have been proposed to explain the transition. However, a comprehensive understanding of the exact origins of the IMT in VO$_2$ still remains elusive. Consequently, this also implies that the exact mechanism of how an external stimulus such as an electric field affects the IMT in VO$_2$ also remains to be completely understood.

*Electronically driven IMT in VO$_2$*: Two-terminal VO$_2$ devices exhibit an IMT when a voltage is applied across the VO$_2$ channel. In this configuration, both electric-field and current-induced Joule heating effects are present[2]. While the exact origin of this transition is also disputed, there is increasing evidence of the preponderance of electro-thermal effects[3,4]. In a three-terminal device, a true gate-field induced IMT has not been demonstrated, although several useful features of the interaction between the (gate) electric-field and the VO$_2$ channel have been revealed (we note that Nakano *et al*[5]. demonstrated a non-volatile phase transition using ionic liquid gating where the role of ionic diffusion and the electric field are challenging to deconvolute). A key feature of the application of an electric field (through the gate) on the VO$_2$ channel is that it modulates the (threshold) voltage required at the source-drain to induce IMT[6] - a property that facilitates the design of the Mott FeFET proposed here. One possible explanation for this behavior is that even though the magnitude of the electric-field required to induce an IMT is significantly larger than that which can supported by a solid state dielectric, the field modulates the nucleation probability of the metallic phase (field induced nucleation) resulting in the change in the threshold voltage. We therefore propose a phenomenological model to emulate this behavior.

## S4. Design and operation of Mott-FeFET array

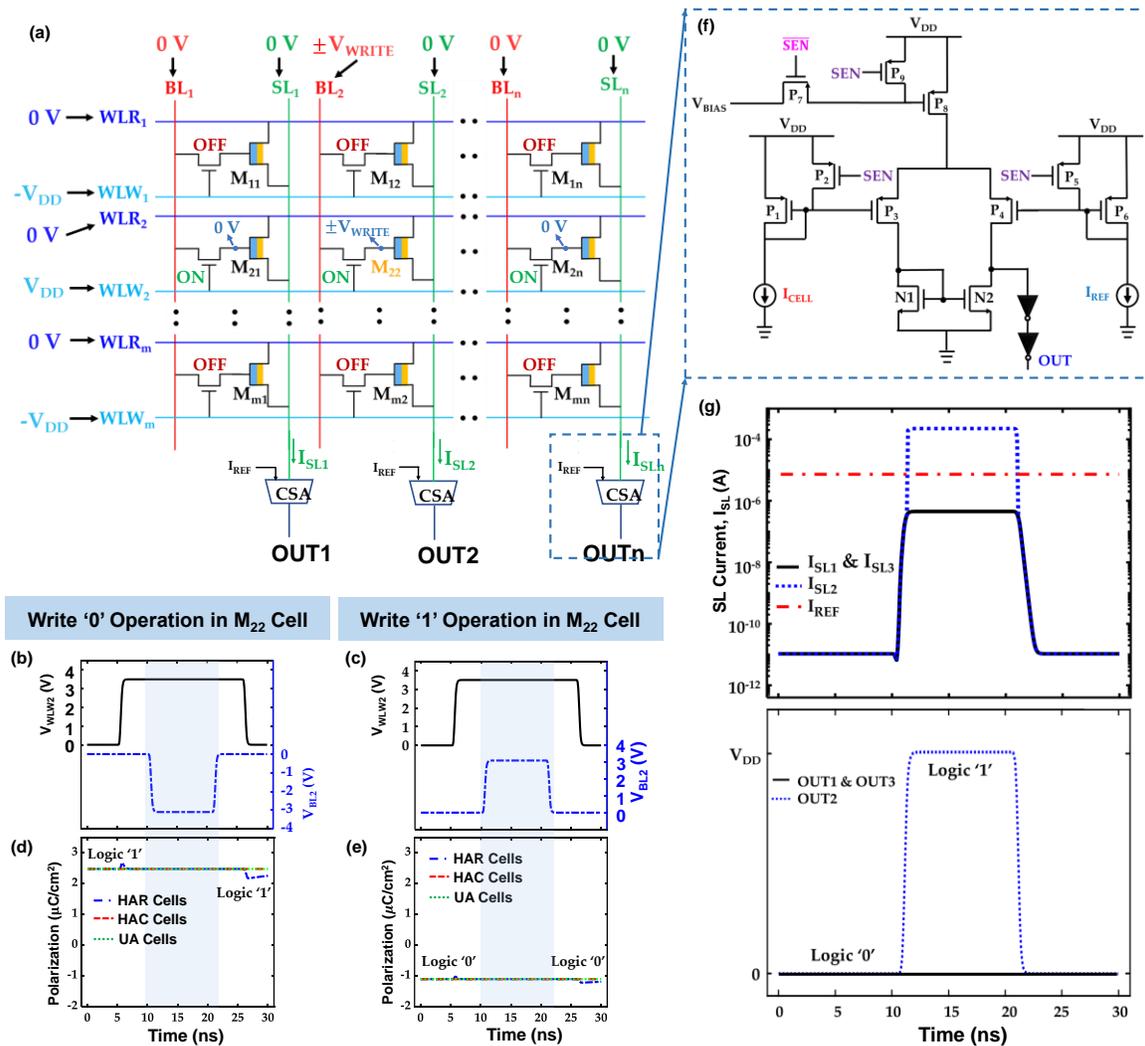

**Fig. S4: Additional Details for the Write Operation and Sensing Mechanism of the Array.** (a) Biasing conditions for WLWs, WLRs, BLs, and SLs during the write operation in $M_{22}$ cell. The choice of biasing for WLWs and SLs ensures that only the access transistors in the same row with the accessed cell turn on while all the other transistors remain off. Now, the choice of biasing for BLs ensures that only the accessed cell ($M_{22}$) gets the write voltage ($\pm V_{WRITE}$) at the gate while all the other cells of that row get 0 V at their gate terminals. A current sense amplifier (CSA) is connected to each SL for the sensing purpose. (b) & (c) Time dynamics of the bias voltages of $WLW_2$ and $BL_2$ during write '0' and write '1' operations in $M_{22}$, respectively. (d) & (e) Preservation of the memory states of HAR, HAC, and UA cells during write '0' and write '1' operations, respectively due to the suitable choice of biasing for the WLWs and BLs. (f) The circuit schematic of the current sense amplifier. (g) SL currents and CSA outputs for read operation in the second row.

Figure 3 of the main text demonstrates the write and read operations in a memory cell within a 3 × 3 array. Additional details are described here. During write/read operations, it is important to ensure the following:

*(i)* The stored data in the inactive cells does not get disturbed during the read/write operations

*(ii)* During the read operation, the SL current shows distinguishable difference for low/high memory states

We first discuss the possibility of the accidental manipulation of the data stored in the inactive cells. Note, in the Mott-FeFET based memory cells, the write and the read operations are performed using the gate and drain bias, respectively. For write (read) operation, suitable $V_{WRITE}$ ($V_{READ}$) is applied as $V_{GS}$ ($V_{DS}$), while $V_{DS}$ ($V_{GS}$) is kept at 0 V. Thus, accidental programming in the inactive cells can be avoided while reading from the active cells. Now, during the write operation, the biasing conditions for the WLWs, WLRs, BLs and SLs (shown in Fig. S4a) are carefully chosen to ensure that, only the accessed cell gets the programming voltage at the gate terminal. The biasing of WLWs (Fig. S4b,c) ensures that the access transistors of half-accessed column (HAC) and unaccessed (UA) cells remain off[7]. Although the access transistors of half-accessed row (HAR) cells turn ON, the BL biasing (Fig. S4b,c) ensures that the HAR cells get 0 V at their gate terminals. Therefore, the memory states of the HAR, HAC, UA cells remain undisturbed during both the write '0' (Fig. S4d), and the write '1' (Fig. S4e) operations.

Next, we discuss the distinguishability in the SL currents during the read operation. We utilize the difference in the SL currents to sense the memory states stored in the cells of an array. Fig. 4h in the main text shows the SL currents for the read operation of the memory cells in the second row. It clearly shows that the SL currents provide sufficient distinguishability between logic '0' and logic '1' states. For sensing, a current sense amplifier (CSA) is connected to each SL, as shown in Fig. S4a. Figure S4b shows the schematic of the CSA[8] that we have used in this work. The reference current ($I_{REF}$= 10 µA) is appropriately chosen to obtain different binary outputs (0 and $V_{DD}$) for logic '0' and logic '1', respectively. Figure S4c shows the SL currents and corresponding logic outputs of the CSA during the read operation.